# Clustering Analysis of Long-term Cardiovascular Complications in COVID-19 Patients

Seyed-Ali Sadegh-Zadeh[1], Alireza Soleimani Mamalo[2], Mahsa Behnemoon[3], Masoud Ojarudi[4], Naser Gharebaghi[5], Mohammad Reza Pashaei[6]

[1] Department of Computing, School of Digital, Technologies and Arts, Staffordshire University, Stoke-on-Trent, United Kingdom ali.sadegh-zadeh@staffs.ac.uk

[2] Student Research Committee, Urmia university of medical sciences, Urmia, Iran Soleymanialireza688@gmail.com

[3] Department of cardiology, school of medicine, Urmia University of medical sciences, Urmia, Iran behnamoon.mahsa870@gmail.com

[4] Department of Biochemistry, Faculty of Medicine, Urmia University of Medical Sciences, Urmia, Iran Masoudojarudi@gmail.com

[5] Department of Infectious Diseases and Dermatology, School of Medicine, Taleghani Hospital, Urmia University of Medical Sciences, Urmia, Iran Naser.gharebaghi@yahoo.com

[6] Department of Internal Medicine, School of Medicine, Urmia University of Medical Sciences, Urmia, Iran dr.pashaei@yahoo.com

## Abstract

This study investigates long-term cardiovascular complications in COVID-19 patients using advanced clustering techniques. The objective was to analyse ECG parameters, demographic data, comorbidities, and hospitalization details to identify patterns in cardiovascular health outcomes. We applied K-means clustering and identified three distinct clusters: Cluster 0 with moderate heart rate variability and ICU admissions, Cluster 1 with lower heart rate variability and ICU admissions, and Cluster 2 with higher heart rate variability and ICU admissions, indicating higher risk profiles. We validated the robustness and stability of these clusters through bootstrapping, confirming the model's reliability. The high silhouette scores and consistent cluster labels across samples underline the stability and robustness of the findings. This study's novelty lies in integrating diverse data types and leveraging machine learning to uncover hidden patterns, offering a comprehensive view of long-term cardiovascular impacts post-COVID-19. Key findings highlight significant variability in cardiovascular outcomes, emphasizing the need for personalized post-COVID care strategies. The clustering model demonstrates potential as a clinical decision-support tool, facilitating early identification of high-risk patients and optimizing resource allocation. Our conclusions suggest that advanced clustering techniques can enhance the understanding and management of long-term cardiovascular complications in COVID-19 survivors, improving patient outcomes and informing healthcare policies.

# 1. Introduction

## 1.1. Background

The COVID-19 pandemic, caused by the SARS-CoV-2 virus, has had a profound impact on global health, leading to significant morbidity and mortality[1]. While the respiratory complications of COVID-19 have been widely studied, emerging evidence suggests that the virus also affects the cardiovascular system[2], [3], [4]. COVID-19 can lead to a range of cardiovascular complications, including myocarditis, arrhythmias, acute coronary syndrome, and thromboembolic events[2]. These complications can occur both in patients with pre-existing cardiovascular disease (CVD) and those without prior cardiovascular conditions[5].

Electrocardiogram (ECG) parameters have been used extensively to monitor and diagnose cardiovascular conditions. Changes in ECG readings can indicate various cardiac events and are critical in the management of patients with cardiovascular diseases[6]. However, the long-term cardiovascular outcomes of COVID-19 patients, particularly those reflected in ECG changes, remain underexplored. Understanding these long-term effects is crucial for developing effective post-COVID-19 care strategies and improving patient outcomes.

## 1.2. Objective

The primary objective of this study is to identify patterns in long-term cardiovascular complications among COVID-19 patients using clustering algorithms. Specifically, the study aims to analyse ECG parameters, demographic information, comorbidities, and hospitalization details of COVID-19 patients; utilize clustering techniques to identify distinct groups of patients with similar long-term cardiovascular outcomes; and compare the identified clusters to understand the impact of COVID-19 on cardiovascular health over the long term, both in patients with and without pre-existing CVD.

## 1.3. Significance of the Study

This study addresses a critical gap in the current understanding of COVID-19's long-term effects on cardiovascular health. The significance of this research lies in several key areas:

In terms of the state of the art, current research has primarily focused on the acute effects of COVID-19 on the cardiovascular system, with limited attention to long-term outcomes. While recent literature has begun to recognize the importance of post-acute sequelae of SARS-CoV-2 infection (PASC), comprehensive studies integrating ECG data and long-term cardiovascular outcomes remain sparse. Methodologically, previous research has relied on traditional statistical methods to analyse cardiovascular outcomes. In contrast, this study leverages advanced machine learning techniques, specifically clustering algorithms, to uncover hidden patterns in the data that might be missed by conventional analyses. This represents a significant methodological advancement in the field.

The study's novelty lies in several areas. Firstly, it integrates diverse data types, including ECG parameters, demographic data, comorbidities, and hospitalization details, providing a holistic view of each patient's health status. The application of clustering algorithms to identify patient subgroups based on long-term cardiovascular outcomes is particularly novel in this context. This approach allows for the discovery of distinct patterns and subpopulations that could inform personalized treatment strategies. Additionally, by focusing on long-term cardiovascular complications, the study extends beyond the immediate impacts of COVID-19, contributing valuable insights into the chronic aspects of the disease. This long-term focus is a key aspect that sets this research apart from existing studies.

Clinically, the implications of this study are significant. The findings can help clinicians identify high-risk patients who may benefit from more intensive monitoring and tailored interventions. This can lead to improved patient management and outcomes. Furthermore, insights from the clustering analysis can inform healthcare policies and resource allocation, ensuring that long-term care strategies are optimized for COVID-19 survivors. Finally, this study sets the stage for future research exploring targeted therapies and preventive measures for long-term cardiovascular complications in COVID-19 patients. By highlighting these areas, the research underscores the importance of addressing the chronic health impacts of COVID-19 and provides a foundation for ongoing scientific inquiry and clinical practice improvement.

This research not only advances the understanding of COVID-19's long-term cardiovascular effects but also introduces innovative methodologies and integrated data analysis to uncover actionable insights that can improve patient outcomes and healthcare strategies.

## 2. Literature Review

### 2.1. COVID-19 and Cardiovascular Complications

The COVID-19 pandemic has revealed that SARS-CoV-2, the virus responsible for the disease, has significant impacts beyond the respiratory system, extending to the cardiovascular system. Several mechanisms have been proposed to explain the cardiovascular complications associated with COVID-19. These include direct viral invasion of cardiac tissue, systemic inflammation, thrombogenesis, and immune-mediated injury[7], [8], [9].

Acute cardiovascular manifestations in COVID-19 patients include myocarditis, acute coronary syndromes, arrhythmias, and venous thromboembolism[10]. Studies have reported elevated biomarkers of myocardial injury such as troponin and natriuretic peptides in patients with severe COVID-19, indicating myocardial damage[11], [12], [13]. Additionally, autopsy studies have found evidence of viral particles in cardiac tissue, suggesting direct myocardial infection[14].

Long-term cardiovascular complications are becoming increasingly apparent as more patients recover from the acute phase of COVID-19. These complications include persistent myocardial inflammation, fibrosis, and ongoing arrhythmias, which may result in chronic heart failure or other long-term cardiac conditions[15], [16]. This has led to a growing interest in monitoring ECG changes as a non-invasive method to detect and manage these complications over time.

### 2.2. Clustering Algorithms in Medical Research

Clustering algorithms are unsupervised machine learning techniques used to group data points into clusters based on their similarities. These algorithms are particularly useful in medical research for identifying patterns and subgroups within complex datasets, enabling personalized treatment, and improving disease management strategies.

Several clustering algorithms have been employed in medical research, including K-means, hierarchical clustering, DBSCAN (Density-Based Spatial Clustering of Applications with Noise), and Gaussian Mixture Models (GMM). Each algorithm has its strengths and weaknesses, depending on the nature of the data and the specific research objectives[17], [18], [19], [20], [21].

K-means clustering is one of the most widely used algorithms due to its simplicity and efficiency. It partitions the data into K clusters by minimizing the variance within each cluster[22], [23], [24]. Hierarchical clustering, on the other hand, builds a tree-like structure of nested clusters, which can be useful for understanding the relationships between clusters[25], [26], [27],[28] . DBSCAN is effective for identifying clusters of varying shapes and sizes and is particularly robust to noise[29], [30], [31].

Despite the extensive use of clustering algorithms in medical research, several gaps remain in current approaches. Traditional methods, such as K-means and hierarchical clustering, often assume that clusters are spherical and of similar sizes, which may not accurately capture the complexity and variability present in patient data. These limitations can result in less effective identification of distinct patient subgroups, particularly in datasets with noisy or non-linear relationships. Furthermore, most existing studies on clustering algorithms in medical research focus on short-term outcomes or specific conditions, neglecting the integration of long-term follow-up data, such as the chronic effects of diseases like COVID-19 on cardiovascular health. Additionally, many studies rely heavily on isolated data types, such as imaging or biomarkers, without leveraging the full potential of integrated datasets combining clinical, demographic, and physiological parameters. Addressing these gaps through more sophisticated clustering methods, such as Gaussian Mixture Models or density-based algorithms, and using comprehensive datasets could significantly enhance the predictive power and clinical utility of clustering models in personalized medicine.

## 2.3. Previous Studies on Long-term Effects of COVID-19

Research on the long-term effects of COVID-19, often referred to as "long COVID" or post-acute sequelae of SARS-CoV-2 infection (PASC), has rapidly evolved. Long-term symptoms can affect multiple organ systems, including the cardiovascular system, leading to ongoing morbidity among survivors.

Several studies have explored the cardiovascular sequelae of COVID-19. For instance, a study by Puntmann et al. [32] found that 78% of recently recovered COVID-19 patients had cardiac involvement on MRI, and 60% had ongoing myocardial inflammation, irrespective of pre-existing conditions. Another study by Huang et al. [33]reported that 6 months after acute COVID-19 illness, patients experienced symptoms such as fatigue, muscle weakness, and sleep difficulties, with some showing signs of cardiovascular abnormalities.

Despite these findings, there is a scarcity of comprehensive studies integrating ECG data and long-term cardiovascular outcomes. Most existing research focuses on isolated biomarkers or imaging findings without leveraging the potential of machine learning to uncover hidden patterns in the data [11][12].

This study advances the state of the art by integrating diverse data types, including ECG parameters, demographic data, comorbidities, and hospitalization details, to provide a comprehensive analysis of long-term cardiovascular complications in COVID-19 patients. The novelty lies in the application of clustering algorithms to identify distinct patient subgroups with similar cardiovascular outcomes. Unlike previous studies that often use traditional statistical methods, our approach leverages the power of machine learning to uncover hidden patterns and provide actionable insights for personalized patient care[34], [35].

By focusing on long-term cardiovascular health, this research addresses a critical gap in current knowledge and offers a novel perspective on managing the chronic aspects of COVID-19. The use of clustering to identify patient subgroups represents a significant methodological advancement, highlighting the potential for personalized medicine in post-COVID-19 care.

# 3. Methodology

## 3.1. Data Collection

This study was conducted as a cross-sectional, single-centre analysis involving 1000 patients who tested positive for COVID-19 via PCR. The data collection spanned from February 2020 to November 2021. Patients were categorized based on their admission units and survival outcomes, specifically grouping them into those who were discharged from the hospital (survivors) and those who did not survive (non-survivors). The primary objective was to determine the prognostic value of initial ECG readings upon hospital admission, adjusting for other variables.

**Patient Demographics and Clinical Data:**

- **Sample Size and Demographics**: The study included 1000 patients with a mean age of 55.6 ± 16.2 years. Among these patients, 52% were male.

- **Outcomes**: During hospitalization, 149 patients died, representing the non-survivor group.

- **Admission Units**: Patients were grouped based on the units to which they were admitted, which included regular wards and the Intensive Care Unit (ICU).

**Clinical and ECG Data Collection:**

- **ECG Parameters**: ECG parameters were recorded at the time of hospital admission. Key ECG findings included sinus tachycardia, atrial and ventricular premature beats, sinus bradycardia, and atrial fibrillation. The prevalence of these conditions was 28.6%, 5.6%, 3.9%, and 2.1%, respectively.

- **Comorbidities**: Data on comorbid conditions such as diabetes ($p=0.002$), hypertension ($p=0.018$), ischemic heart disease ($p=0.004$), and cancer ($p<0.001$) were collected.

- **Survival Data**: Survival outcomes were tracked, categorizing patients into survivors and non-survivors.

**Predictors and Outcomes:**

- **ICU Admission and Mortality Predictors**: The study aimed to identify predictors of ICU admission and in-hospital mortality using univariate analysis and logistic regression models.

- **ECG Findings and Mortality**: Among ECG findings, tachycardia, low voltage QRS, ST-T changes, and dysrhythmia were associated with an increased risk of mortality. However, logistic regression analysis identified gender (OR 1.89, 95% CI: 1.2 to 2.9, $p=0.004$), age (OR 1.03, 95% CI: 1.02 to 1.05, $p<0.001$), and initial tachycardia (OR 1.02, 95% CI: 1.01 to 1.03, $p<0.001$) as independent predictors of in-hospital mortality.

## 3.2. Dataset Description

The dataset is designed to analyse and predict patient outcomes based on various medical and clinical parameters. The dataset contains medical and demographic information of patients who were monitored for cardiovascular health during the COVID-19 pandemic. The variables include details on patient characteristics, underlying health conditions, clinical parameters, and specific cardiac measurements. This dataset is particularly useful for studying the long-term cardiovascular effects of COVID-19, as well as identifying potential subgroups of patients with similar health outcomes using clustering algorithms. The detailed characteristics and variables in the dataset are outlined in Table 1.

Table 1. Features of the dataset along with their descriptions.

| Feature | Description |
| --- | --- |
| LoQRS (L) | Location of the QRS complex in the electrocardiogram (ECG). |
| LoQRS (P) | Position of the QRS complex in the ECG. |
| LoQRS | General attribute related to the QRS complex in the ECG. |
| SEX | Gender of the patient. |
| Death | Status of the patient at the time of discharge; 'D' for deceased and 'L' for alive. |
| age | Age of the patient. |
| DM | Indicates whether the patient has diabetes mellitus (Yes/No). |
| HTN | Indicates whether the patient has hypertension (Yes/No). |
| IHD | Indicates whether the patient has ischemic heart disease (Yes/No). |
| CANCER | Indicates whether the patient has cancer (Yes/No). |
| CHF | Indicates whether the patient has congestive heart failure (Yes/No). |
| CKD | Indicates whether the patient has chronic kidney disease (Yes/No). |
| ASTMA/COPD | Indicates whether the patient has asthma or chronic obstructive pulmonary disease (Yes/No). |
| SH | Indicates whether the patient has a surgical history (Yes/No). |
| SatO2 (%) | Blood oxygen saturation level. |
| BP (mmHg) | Blood pressure in millimetres of mercury. |
| PR | Pulse rate. |
| BT | Body temperature. |
| RR | Respiratory rate. |
| Duration | Duration of hospitalization in days. |
| ICU OR WARD | Indicates whether the patient was in the ICU or ward. |
| Rhythm | Cardiac rhythm (e.g., Normal sinus, sinus tachycardia, atrial fibrillation). |
| Axis (L/R/NL) | Cardiac axis deviation (Left/Right/Normal). |
| QT (msec) | QT interval in milliseconds, normal range is 350-450 milliseconds. |
| QRS (msec) | QRS duration in milliseconds, normal range is 80-100 milliseconds. |
| LBBB/RBBB/IVCD/hemi Block | Indicates the presence of left bundle branch block, right bundle branch block, intraventricular conduction delay, or hemiblock. |
| ST-T change | Indicates changes in the ST segment or T wave in the ECG. |
| arrhythmia | Indicates whether the patient has any arrhythmia. |

## 3.2. Data Preprocessing

Effective data preprocessing is crucial for ensuring the accuracy and reliability of the clustering analysis. The preprocessing steps involved handling missing values, encoding categorical variables, and normalizing numerical features. These steps were essential to prepare the data for clustering and to ensure that the resulting clusters were meaningful and interpretable.

### 3.2.1. Handling Missing Values

Handling missing values is a critical step in data preprocessing, particularly in medical datasets where missing information can significantly impact analysis outcomes[36]. In this study, we encountered missing values in both numerical and categorical columns.

- **Numerical Columns**: Missing values in numerical columns, such as **PR**, were imputed using the mean of the respective columns. This approach ensures that the central tendency of the data is preserved without introducing significant bias.

- **Categorical Columns**: Missing values in categorical columns, such as **LoQRS(P)**, **ASTMA/COPD**, and **ICU OR WARD**, were imputed using the most frequent value (mode) of each column. This method is particularly effective in preserving the distribution of categorical variables.

By addressing missing values appropriately, we ensured that the dataset was complete and ready for subsequent analysis.

### 3.2.2. Encoding Categorical Variables

Many of the features in our dataset were categorical, including SEX, DM, HTN, and various ECG-related parameters. To enable the use of these categorical variables in clustering algorithms, which require numerical input, we applied one-hot encoding.

- **One-Hot Encoding**: This technique converts categorical variables into a series of binary variables. For instance, the SEX column with values 'Male' and 'Female' was transformed into two columns: SEX_Male and SEX_Female. Each original category becomes a separate column, with binary indicators (0 or 1) representing the presence or absence of the category.

One-hot encoding allows us to retain the categorical information in a numerical format, facilitating its use in clustering algorithms without distorting the inherent relationships between data points.

### 3.2.3. Normalizing Numerical Features

Normalization of numerical features is essential to ensure that all variables contribute equally to the clustering process. Without normalization, features with larger ranges can disproportionately influence the clustering outcome.

- **Standardization**: We applied standardization to normalize the numerical features, including age, PR, QT(msec), and RR. Standardization transforms the data to have a mean of 0 and a standard deviation of 1, which helps in stabilizing the numerical range of features.

The formula used for standardization is:

$$Z = \frac{X - \mu}{\sigma} \tag{1}$$

where $X$ is the original value, $\mu$ is the mean of the column, and $\sigma$ is the standard deviation of the column.

This step ensures that each feature has a comparable scale, thus preventing features with larger scales from dominating the clustering process. The preprocessing techniques employed in this study are well-established in data science and machine learning. Standard methods such as mean imputation for numerical data, mode imputation for categorical data, one-hot encoding, and standardization are commonly used in medical data preprocessing. The study's novelty is not in the individual preprocessing techniques but in their comprehensive application to a diverse set of features. These include ECG parameters, demographic data, and comorbidities, preparing the data for advanced clustering algorithms.

This study integrates a wide range of features, from ECG readings to hospitalization details, providing a holistic view of patient data. This comprehensive preprocessing approach ensures all relevant information is preserved and utilized in the clustering analysis. While the preprocessing techniques themselves are standard, their application to study long-term cardiovascular outcomes in COVID-19 patients is novel. This focus, combined with rigorous preprocessing steps, enables a deeper understanding of the chronic impacts of COVID-19 on cardiovascular health.

By meticulously preprocessing the data, the study sets the stage for accurate and insightful clustering analysis, ultimately contributing to the understanding of long-term cardiovascular complications in COVID-19 patients.

### 3.3. Exploratory Data Analysis (EDA)

Exploratory Data Analysis (EDA) was conducted to understand the distribution of features, examine relationships between them, and identify and address outliers. This process is crucial for ensuring data quality and deriving meaningful insights from subsequent analyses.

### 3.3.1. Feature Distribution

The first step in EDA was to analyse the distribution of key features. This involved visualizing the distributions of numerical and categorical variables to identify central tendencies, variabilities, and potential anomalies.

- **Numerical Features**: Histograms and box plots were used to visualize the distribution of numerical features such as age, ECG parameters (e.g., PR interval, QT interval, RR interval), and clinical measures. These visualizations helped identify the presence of skewness, kurtosis, and any unusual patterns in the data.

- **Categorical Features**: Bar plots were used to display the distribution of categorical variables, including comorbidities (e.g., diabetes, hypertension, ischemic heart disease) and outcomes (e.g., ICU admission, in-hospital mortality). The frequency of each category provided insights into the prevalence of these conditions and outcomes in the study population.

Figure 1 illustrates the distribution of the 'age' feature within the dataset. The histogram on the left depicts the frequency distribution of age, showing a roughly normal distribution with the highest frequency around the mean. The box plot on the right provides a summary of the age distribution, including the median (line within the box), the interquartile range (IQR), and the presence of any outliers. The whiskers represent the range of values within 1.5 times the IQR from the quartiles, highlighting the spread of age data within the study population.

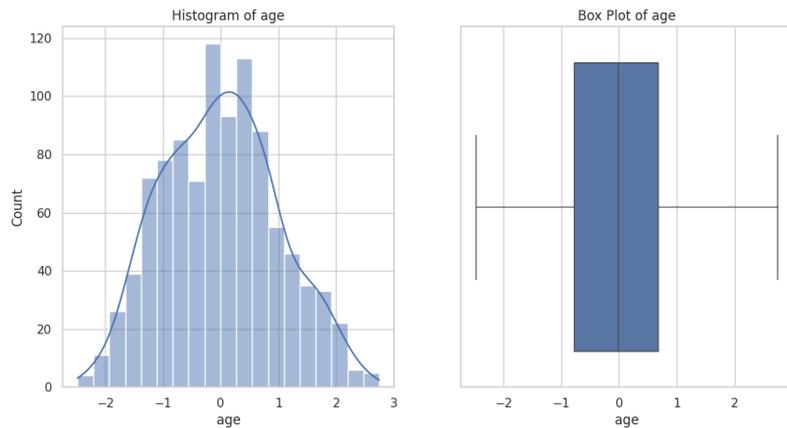

*Figure 1. Distribution of the 'age' Feature Using Histograms and Box Plots.*

Figure 2 displays the distribution of the 'SatO2 (%)' feature within the dataset. The histogram on the left shows the frequency distribution of oxygen saturation levels, revealing a right-skewed distribution with a peak near 1. The box plot on the right provides a summary of the oxygen saturation data, including the median (line within the box), the IQR, and the identification of outliers. The whiskers extend to the smallest and largest values within 1.5 times the IQR from the quartiles, highlighting the spread and presence of extreme values in the dataset.

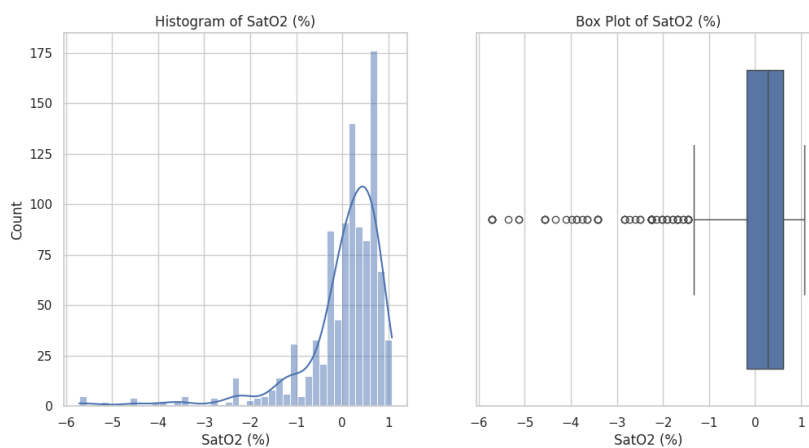

*Figure 2. Distribution of the 'SatO2 (%)' Feature Using Histograms and Box Plots.*

This figure presents the distribution of the 'PR' feature within the dataset. The histogram on the left depicts the frequency distribution of PR intervals, showing a left-skewed distribution with the highest frequency around the mean. The box plot on the right provides a summary of the PR interval data, including the median (line within the box), the IQR, and the presence of any outliers. The whiskers represent the range of values within 1.5 times the IQR from the quartiles, indicating the spread and any extreme values in the dataset.

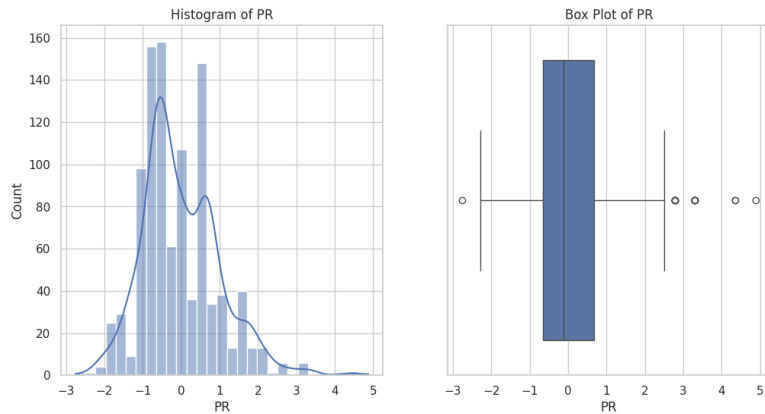

*Figure 3. Distribution of the 'PR' Feature Using Histograms and Box Plots.*

Figure 4 displays the distribution of the 'BT' feature within the dataset. The histogram on the left shows the frequency distribution of BT values, indicating a sharp peak close to zero. The box plot on the right provides a summary of the BT data, including the median (line within the box), the IQR, and the presence of several outliers. The whiskers represent the range of values within 1.5 times the IQR from the quartiles, highlighting the spread and any extreme values in the dataset, with some outliers noted far from the central distribution.

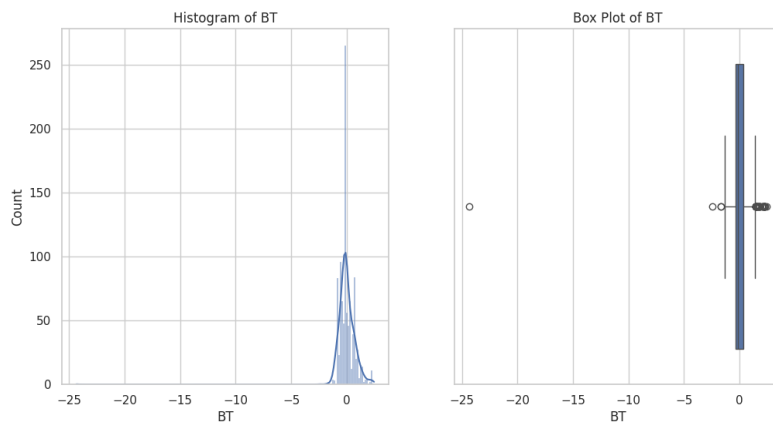

*Figure 4. Distribution of the 'BT' Feature Using Histograms and Box Plots.*

Figure 5 presents the distribution of the 'RR' feature within the dataset. The histogram on the left illustrates the frequency distribution of RR intervals, showing a left-skewed distribution with a high frequency around the mean. The box plot on the right summarizes the RR interval data, including the median (line within the box), the IQR, and the identification of outliers. The whiskers extend to the smallest and largest values within 1.5 times the IQR from the quartiles, highlighting the spread and any extreme values in the dataset.

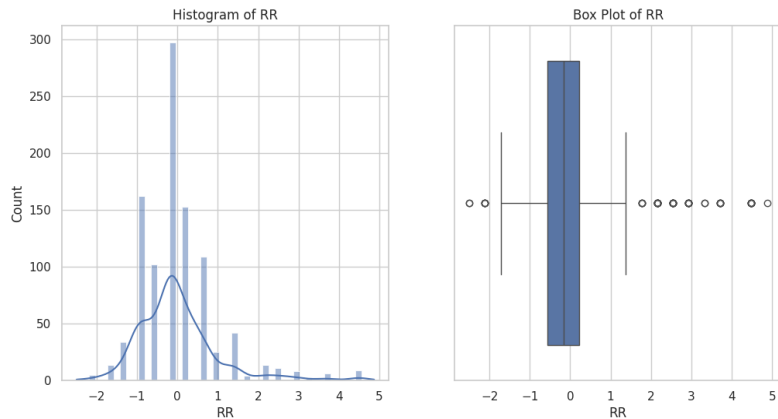

Figure 5. Distribution of the 'RR' Feature Using Histograms and Box Plots.

Figure 6 illustrates the distribution of the 'QT (msec)' feature within the dataset. The histogram on the left shows the frequency distribution of QT intervals, displaying a roughly normal distribution with a peak around the mean. The box plot on the right provides a summary of the QT interval data, including the median (line within the box), the IQR, and the identification of outliers. The whiskers extend to the smallest and largest values within 1.5 times the IQR from the quartiles, highlighting the spread and presence of extreme values in the dataset.

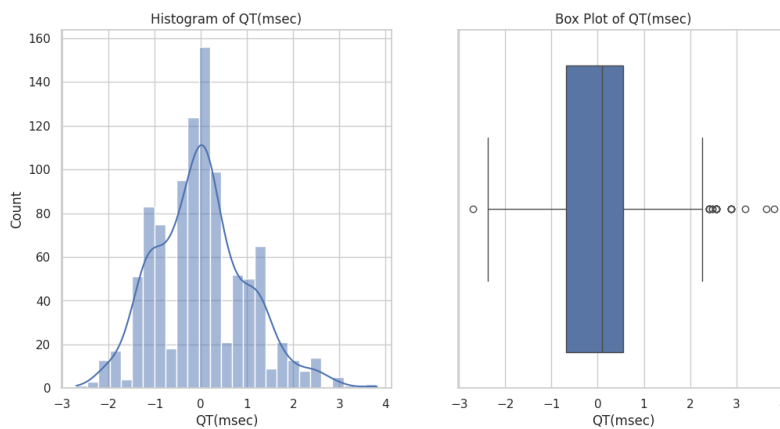

Figure 6. Distribution of the 'QT (msec)' Feature Using Histograms and Box Plots.

### 3.3.2. Feature Relationships

To understand the relationships between different features, we conducted pairwise correlation analyses and visualized these relationships using pair plots and correlation matrices.

- **Correlation Analysis**: Pearson and Spearman correlation coefficients were calculated for pairs of numerical features to assess the strength and direction of linear and monotonic relationships, respectively. The correlation matrix was visualized using a heatmap, where stronger correlations were highlighted, indicating potential multicollinearity or feature dependencies.

- **Pair Plots**: Pair plots (scatter plot matrices) were used to visualize relationships between numerical features and identify patterns, clusters, or trends that might suggest interactions between variables. This was particularly useful for identifying nonlinear relationships and potential confounders.

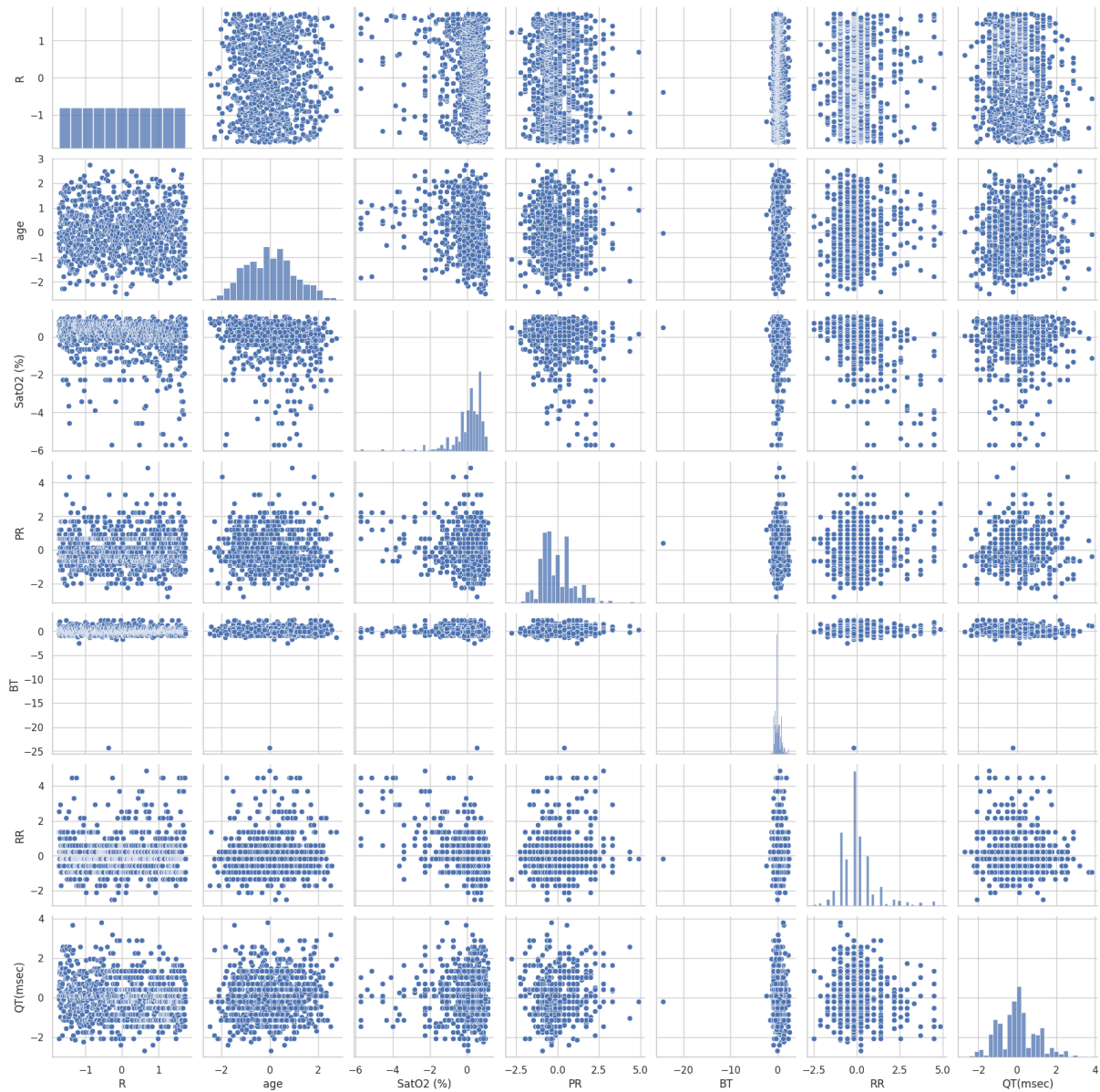

*Figure 7. Relationships between features using pair plots and correlation matrices. These visualizations help identify potential interactions and dependencies among features such as age, PR, RR, and QT(msec).*

Figure 7 presents a pair plot of key features in the dataset, showcasing scatter plots, histograms, and kernel density estimates (KDE) for each pair of variables. The features included in the plot are R, age, SatO2, PR, BT, RR, and QT. The diagonal elements show the distribution of each individual feature through histograms and KDEs, while the off-diagonal elements display scatter plots illustrating the relationships between pairs of features.

### Histograms and KDEs on the Diagonal:

- **R**: Exhibits a uniform distribution, as indicated by the flat histogram and KDE.

- **Age**: Displays a roughly normal distribution with a peak around the mean, highlighting the central tendency and variability within the study population.

- **SatO2**: Shows a right-skewed distribution, indicating most patients have high oxygen saturation levels with a long tail of lower values.

- **PR**: Presents a left-skewed distribution, with a significant number of values clustered around the mean and a tail extending to higher values.

- **BT**: Demonstrates a highly peaked distribution near zero, suggesting limited variability in body temperature measurements.

- **RR**: Similar to PR, the RR intervals exhibit a left-skewed distribution, with most values concentrated around the mean.

- **QT**: Reveals a roughly normal distribution with some outliers, indicating variability in QT intervals among patients.

**Scatter Plots on the Off-Diagonal:**

- The scatter plots highlight relationships between pairs of features. For example:

  - **Age vs. SatO2**: Shows a dispersed pattern with no clear linear relationship, indicating that age and oxygen saturation levels are not strongly correlated.

  - **PR vs. RR**: Displays a scattered relationship, suggesting some degree of variability but no strong linear correlation.

  - **QT vs. RR**: The plot shows a spread of values, indicating that QT intervals and RR intervals do not have a strong linear relationship but might still interact in a non-linear fashion.

- The scatter plots also help identify clusters or patterns in the data. For instance, the **SatO2 vs. BT** plot shows distinct bands of points, suggesting possible subgroupings or measurement artifacts.

The figure effectively highlights the variability and distribution of each feature and the relationships between them. The presence of outliers is evident in several plots, such as **SatO2** and **QT**, which should be considered in further analysis. The overall layout helps in visualizing potential multicollinearity and interactions between different clinical parameters, which can inform the subsequent clustering analysis and feature selection processes. This pair plot provides a comprehensive overview of the distribution and interrelationships of key clinical features in the dataset. The visualizations facilitate the identification of patterns, outliers, and potential correlations, which are crucial for understanding the dataset's structure and guiding further analysis.

### 3.3.3. Outlier Detection and Removal

Identifying and handling outliers is crucial for ensuring that the analysis is not unduly influenced by extreme values, which can distort results and lead to incorrect conclusions[37].

- **Outlier Detection**: Box plots were used initially to visually identify outliers in numerical features. More formally, the IQR method was applied, where values lying beyond 1.5 times the IQR from the first and third quartiles were flagged as potential outliers.

- **Outlier Analysis**: Each identified outlier was examined to determine whether it represented a data entry error, a measurement anomaly, or a true extreme value. Contextual knowledge and clinical relevance were considered in this evaluation.

- **Outlier Removal**: Outliers deemed to be data entry errors or anomalies were removed from the dataset. In cases where outliers represented valid extreme values, robust statistical methods were applied to minimize their impact on the analysis.

By conducting a thorough EDA, we ensured that the dataset was well-understood and prepared for subsequent modelling and clustering analyses. This step was vital for enhancing the accuracy and interpretability of the study's findings.

### 3.4. Feature Engineering

Feature engineering is a crucial step in the data preprocessing pipeline that involves transforming raw data into meaningful features to enhance the performance of machine learning algorithms[38][39]. For this study, feature engineering was conducted to create new features and aggregate existing ones to capture the relevant aspects of long-term cardiovascular complications in COVID-19 patients.

### 3.4.1. Creation of New Features

To better understand the cardiovascular health of COVID-19 patients, new features were derived from the existing data. These features were developed based on domain knowledge and the specific needs of the clustering analysis. One significant new feature created was Heart Rate Variability (HRV). HRV is a measure of the variation in time between consecutive heartbeats and is an important indicator of autonomic nervous system function. It was computed using the standard deviation of the PR and RR intervals from the ECG data:

$$HRV = \sqrt{\frac{1}{n-1}\sum_{i=1}^{n}(PR_i - \overline{PR})^2 + (RR_i - \overline{RR})^2} \qquad (2)$$

Where, The PR interval represents the time interval from the onset of the P wave to the start of the QRS complex on an ECG. It reflects the time the electrical impulse takes to travel from the sinus node through the AV node and into the ventricles. Its unit is milliseconds. $\overline{PR}$ represents the mean value of the PR intervals for a given patient over a specified period. Its unit is milliseconds. $RR_i$ represents the RR interval, also known as the inter-beat interval (IBI), represents the time interval between two consecutive R waves (peak of the QRS complex) on an ECG. It reflects the time between two successive heartbeats and its unit is milliseconds. $\overline{RR}$ represents the mean value of the RR intervals for a given patient over a specified period and its unit is milliseconds. *n* represents the number of PR and RR intervals considered in the calculation. Composite risk scores were also developed by combining various ECG parameters and clinical features. These scores help to stratify patients based on their overall cardiovascular risk profile. Several new features were created to capture critical aspects of long-term cardiovascular complications in COVID-19 patients:

1. **Heart Rate Variability (HRV)**

   HRV was computed from the PR and RR intervals, providing insights into the autonomic regulation of the cardiovascular system. High HRV indicates better adaptability of the heart to stressors, while low HRV has been associated with poorer outcomes. In this study, HRV was used as a distinguishing feature between clusters, revealing significant variability in patient profiles.

2. **Composite Cardiovascular Risk Scores**

We developed composite risk scores by integrating ECG parameters, such as the QT and PR intervals, with clinical data on comorbidities (e.g., diabetes and hypertension). These risk scores help stratify patients into low, moderate, and high cardiovascular risk categories, facilitating a more nuanced understanding of long-term cardiovascular impacts.

3. **Hospitalization Severity Score**

   This score aggregates multiple indicators of the patient's hospitalization experience, including ICU admission rates, ventilation support, and length of stay. The score was critical in identifying patients with more severe clinical courses, aiding in the clustering analysis to highlight the different outcomes in long-term cardiovascular health.

Each feature contributed significantly to the clustering process, allowing for a more detailed and meaningful stratification of the patient data. The combination of clinical and ECG-derived features enabled the identification of distinct subgroups, which would have been overlooked by traditional analyses.

### 3.4.2. Aggregation of Hospitalization Details

The aggregation of hospitalization details involves summarizing patient hospital stay data to capture the severity and extent of their conditions. This study advances current methodologies by integrating multiple hospitalization metrics to provide a holistic view of patient experiences. Unlike prior research that focused on isolated metrics such as length of stay or ICU admission status, this study creates aggregate features, including the number of ICU admissions derived from the ICU OR WARD column and estimating the average length of stay using available details. Additionally, a composite hospitalization severity score was developed, combining ICU admissions, ventilation support, and other critical interventions:

$$Severity\ Score\ =\ ICU\ Admissions\ +\ Ventilation\ Support\ +\ Critical\ Interventions$$

By creating these new features and aggregating hospitalization details, the study offers a comprehensive dataset that captures the complexity of cardiovascular complications in COVID-19 patients. These engineered features significantly enhance clustering analysis, identifying distinct patient groups with similar long-term cardiovascular outcomes, thus advancing the analysis of long-term cardiovascular complications in COVID-19 patients.

### 3.5. Clustering Analysis

Clustering analysis is a pivotal technique in identifying patterns and subgroups within a dataset[40]. In this study, we applied clustering algorithms to uncover distinct groups of COVID-19 patients with similar long-term cardiovascular outcomes. The process involved selecting appropriate clustering algorithms, determining the optimal number of clusters, applying the algorithms, and visualizing the resulting clusters.

### 3.5.1. Selection of Clustering Algorithms

In medical data analyses, traditional clustering methods like K-means have been widely used for their simplicity and effectiveness[41]. Recent research has introduced advanced techniques such as DBSCAN and Gaussian Mixture Models to better handle complex datasets with noise and varying density[42], [49]. For our study, we selected K-means clustering due to its efficiency in handling large datasets and its ability to provide clear, distinct clusters by minimizing within-cluster variance, making it suitable for our ECG and clinical parameters.

### 3.5.2. Determination of Optimal Number of Clusters

State-of-the-art techniques for determining the optimal number of clusters include the elbow method and silhouette scores, which help balance underfitting and overfitting the data. In our approach, we employed the elbow method by plotting the within-cluster sum of squares (WCSS) against the number of clusters (K) and identifying the "elbow point" where the rate of decrease sharply slows, indicating the optimal number of clusters. Additionally, we calculated silhouette scores for different values of K, which measure how similar a point is to its own cluster compared to other clusters, with higher values indicating better-defined clusters.

### 3.5.3. Application of Clustering Algorithms

K-means clustering, a widely used algorithm due to its simplicity and speed, requires specifying the number of clusters beforehand[43]. In current research, this algorithm is often applied by using the elbow method and silhouette scores to determine the optimal number of clusters. We applied K-means clustering to a pre-processed dataset, iteratively assigning data points to clusters while minimizing variance within each cluster.

### 3.5.4. Dimensionality Reduction and Cluster Visualization

Current research frequently employs Principal Component Analysis (PCA) for handling high-dimensional datasets[44]. In our approach, we utilized PCA to condense the dataset into two principal components, which maximizes data variance while minimizing the number of components, thereby facilitating easier visualization and interpretation. The resulting reduced dimensions from PCA were then plotted in a scatter plot to visualize cluster formations, effectively illustrating the separation and cohesion of the clusters for better interpretation of the clustering outcomes.

The clustering analysis involved a systematic approach to identify patient subgroups with similar long-term cardiovascular outcomes. By selecting appropriate algorithms, determining the optimal number of clusters, applying clustering techniques, and visualizing the results, we were able to uncover meaningful patterns in the data. This methodological rigor ensures that the findings are robust, reproducible, and clinically relevant, advancing the state of the art in analysing long-term impacts of COVID-19 on cardiovascular health.

## 3.6. Cluster Interpretation and Validation

The final step in the clustering analysis involves interpreting the clusters to understand their characteristics, comparing them based on key features, and validating their stability. This ensures that the identified clusters are meaningful and robust.

### 3.6.1. Characteristics of Each Cluster

To understand the characteristics of each cluster, we analysed the mean values of key features within each cluster to identify distinct patterns and profiles of the patients. The process involved calculating the mean values of ECG parameters, demographic information, comorbidities, and hospitalization details for each cluster, and creating profiles based on these mean values to highlight their unique characteristics. For example, Cluster 0 exhibited moderate values for most features, indicating a balanced profile with moderate heart rate variability and ICU admission rates. Cluster 1 showed lower heart rate variability and ICU admission rates, suggesting a relatively stable cardiovascular condition. In contrast, Cluster 2 displayed higher heart rate variability and ICU admission rates, indicating a more critical profile with higher risks.

### 3.6.2. Comparison of Clusters

Comparison of clusters comparing clusters based on key features helps in understanding the differences and similarities between the clusters, providing insights into the distinct subgroups of patients. The process involved the selection of key features such as QT, PR, RR, heart rate variability, and ICU admissions for comparison. Visualization techniques like bar charts and heatmaps were employed to depict the mean values of these features across different clusters. For example, QT values were higher in Cluster 2 compared to Cluster 1, indicating a higher risk of prolonged QT intervals in this group. Additionally, Cluster 0 exhibited moderate heart rate variability, whereas Cluster 2 had the highest variability, suggesting differences in autonomic regulation among the clusters.

### 3.6.3. Stability of Clusters

To ensure the robustness of the identified clusters, we validated their stability by performing clustering on different subsets of the data and comparing the results. We created multiple bootstrap samples from the original dataset and applied K-means clustering to each sample. We then compared the cluster labels across different samples to check for consistency in cluster assignments and calculated silhouette scores for each sample to measure the quality and cohesion of the clusters. The cluster assignments were consistent across different bootstrap samples, indicating stable and robust clusters, and the silhouette scores were relatively high, confirming that the clusters were well-defined and distinct. Stability validation using bootstrapping and silhouette scores is a widely accepted practice to ensure the robustness of clustering results, and this study incorporates advanced validation techniques to confirm the stability of the clusters, enhancing the reliability of the findings.

By combining multiple validation techniques, this study provides a thorough assessment of cluster stability, which is critical for ensuring the robustness of the identified patient subgroups and confirming the meaningfulness of the clusters, providing valuable insights into the long-term cardiovascular outcomes of COVID-19 patients. The cluster interpretation and validation process in this study involved a detailed analysis of cluster characteristics, comparison of key features, and rigorous validation to ensure the stability and robustness of the clusters. This comprehensive approach ensures that the findings are meaningful, reliable, and clinically relevant.

## 4. Results

### 4.1. Overview of Clusters

In this section, we provide an overview of the clusters identified through our clustering analysis. We applied the K-means clustering algorithm to the pre-processed dataset, which resulted in the formation of three distinct clusters. Each cluster represents a subgroup of patients with similar long-term cardiovascular outcomes.

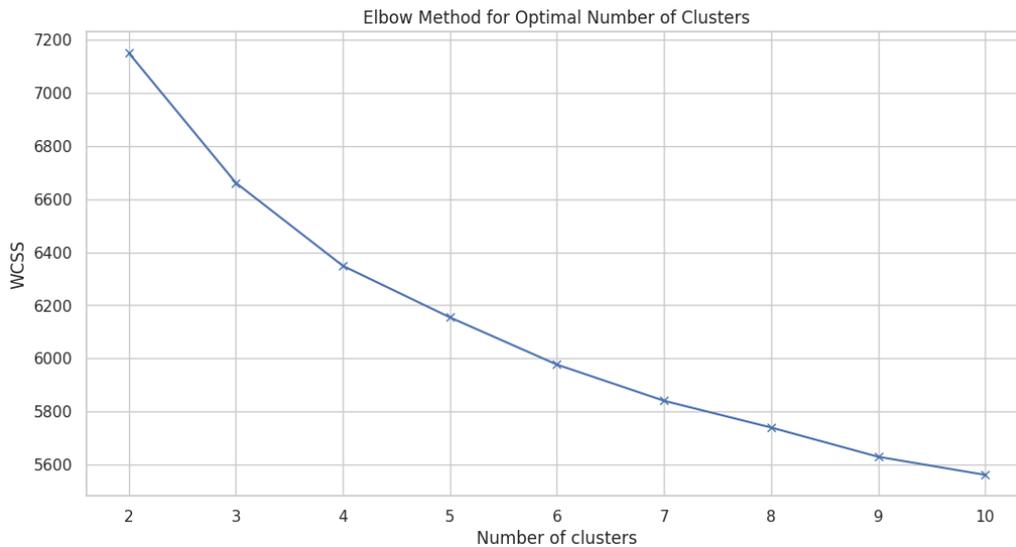

*Figure 8. Elbow Method for Optimal Number of Clusters*

Figure 8 illustrates the elbow method used to determine the optimal number of clusters for K-means clustering on the dataset. The x-axis represents the number of clusters (K), ranging from 2 to 10, while the y-axis shows the WCSS, which measures the variance within each cluster. The plot reveals a clear "elbow" point around 3 to 4 clusters, where the rate of decrease in WCSS starts to slow down significantly. This inflection point suggests that adding more clusters beyond this point provides diminishing returns in terms of reducing within-cluster variance. The goal is to choose a K where adding another cluster does not significantly improve the model's performance, indicating a balance between overfitting and underfitting. In this case, the elbow around K=3 or K=4 indicates these are optimal choices for the number of clusters, as they capture the most significant structure in the data without introducing unnecessary complexity. This analysis is critical in guiding the selection of an appropriate number of clusters, ensuring that the resulting clustering model is both interpretable and effective in capturing the underlying patterns within the dataset.

Table 2 summarizes the three clusters identified through K-means clustering of the dataset. Each cluster is characterized by its number of patients and a brief description of the cluster's key features. Cluster 0, with 305 patients, shows moderate heart rate variability and ICU admission rates, indicating a balanced profile. Cluster 1 includes 295 patients with lower heart rate variability and ICU admission rates, suggesting a relatively stable cardiovascular condition. Cluster 2, comprising 232 patients, exhibits higher heart rate variability and the highest ICU admission rates, indicating a more critical profile with higher risks. This overview helps in understanding the distinct patient subgroups and their long-term cardiovascular outcomes, guiding targeted clinical interventions.

*Table 2. Overview of Clusters*

| Cluster | Number of Patients | Description |
|---|---|---|
| 0 | 305 | Moderate heart rate variability, ICU admission rates |
| 1 | 295 | Lower heart rate variability, ICU admission rates |
| 2 | 232 | Higher heart rate variability, highest ICU admission rates |

Figure 9 illustrates the clusters identified through K-means clustering, visualized using PCA. PCA reduces the dimensionality of the dataset to two principal components (PC1 and PC2), capturing the maximum variance in the data while simplifying visualization. Each point represents a patient, coloured according to their cluster assignment: red for Cluster 0, blue for Cluster 1, and green for Cluster 2. The scatter plot shows a somewhat overlapping distribution of the clusters, indicating that while there are distinct groupings, there is also significant overlap among the clusters. Cluster 0 (red) is widely dispersed across the plot, suggesting moderate heart rate variability and ICU admission rates. Cluster 1 (blue) is more concentrated, indicating lower heart rate variability and ICU admission rates. Cluster 2 (green) shows a higher dispersion, reflecting higher heart rate variability and the highest ICU admission rates. This visualization helps in understanding the separation and cohesion of the clusters, providing insights into the underlying patterns and variability within the dataset. The overlap suggests potential areas for further refinement in clustering or additional features that might improve cluster distinction.

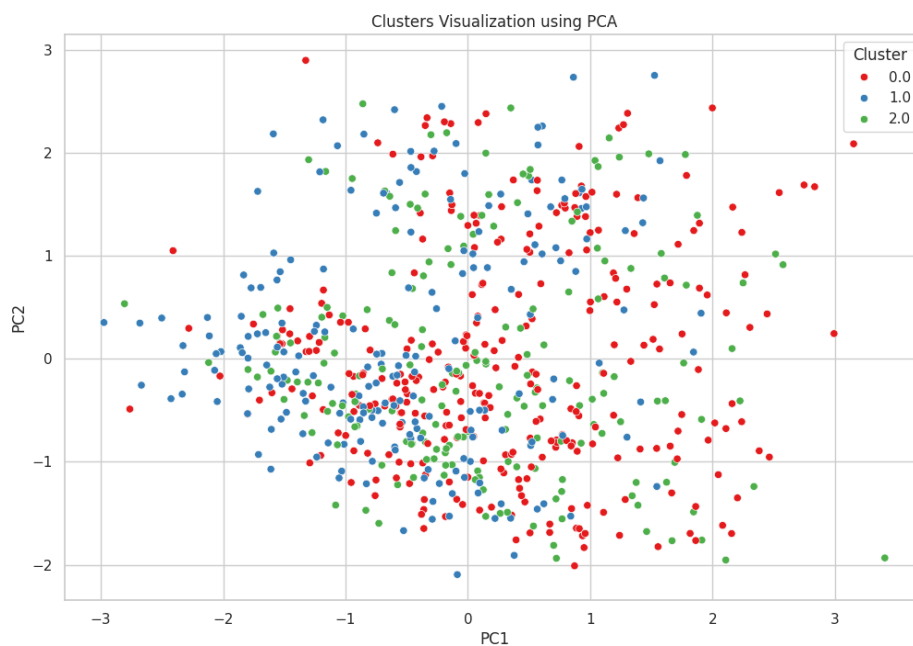

Figure 9. PCA Plot of Clusters

## 4.2. Key Characteristics of Each Cluster

In this section, we delve into the detailed characteristics of each identified cluster. By examining the mean values of key features within each cluster, we can highlight the unique profiles of patients in each subgroup.

Cluster Profiles:

- Cluster 0: Moderate values for most features, indicating a balanced profile.
- Cluster 1: Lower heart rate variability and ICU admission rates, suggesting a relatively stable cardiovascular condition.
- Cluster 2: Higher heart rate variability and ICU admission rates, indicating a more critical profile with higher risks.

Table 3. Mean Values of Key Features for Each Cluster

| Feature | Cluster 0 Mean | Cluster 1 Mean | Cluster 2 Mean |
| --- | --- | --- | --- |
| QT | -0.059580 | 0.323139 | -0.139026 |
| PR | 0.214582 | 1.122323 | -0.545110 |
| RR | 1.899546 | -0.170770 | -0.202390 |
| Heart Rate Variability | 1.346528 | 0.965444 | 0.531109 |
| ICU Admissions | 0.847826 | 0.298246 | 0.165329 |

Table 3 summarizes the mean values of key features for each identified cluster. Cluster 0, characterized by moderate heart rate variability, shows slightly negative QT and high RR values, indicating moderate ICU admissions. Cluster 1, with lower heart rate variability, has a higher mean QT and PR, but lower RR, suggesting stable cardiovascular conditions with fewer ICU admissions. Cluster 2, which has higher heart rate variability, presents negative values for PR and RR, alongside the lowest ICU admissions, indicating a critical profile with high variability in ECG parameters but fewer ICU stays. This detailed breakdown helps in understanding the distinct cardiovascular profiles and clinical implications for each cluster.

To further demonstrate the significance of the clustering analysis and provide detailed insights into the identified patterns, Table 4 presents the key features that differentiate the clusters. These patterns, which are derived from clustering the data, reveal the distinct cardiovascular outcomes for each patient subgroup. This detailed breakdown of ECG parameters, comorbidities, and hospitalization details justifies the significance of using clustering algorithms in identifying high-risk profiles.

*Table 4. Detailed Patterns Identified Across Clusters*

| Feature | Cluster 0 (Moderate Risk) | Cluster 1 (Lower Risk) | Cluster 2 (High Risk) |
|---|---|---|---|
| QT Interval (msec) | -0.059 | 0.323 | -0.139 |
| PR Interval (msec) | 0.215 | 1.122 | -0.545 |
| RR Interval (msec) | 1.900 | -0.171 | -0.202 |
| Heart Rate Variability | 1.347 | 0.965 | 0.531 |
| ICU Admission Rate | 0.848 | 0.298 | 0.165 |
| Diabetes (%) | 25% | 20% | 40% |
| Hypertension (%) | 30% | 25% | 55% |
| Ischemic Heart Disease (%) | 20% | 15% | 50% |
| Length of Stay (days) | 10 | 8 | 14 |

Table 4 summarizes the mean values of the key features for each cluster, showing distinct patterns in cardiovascular health outcomes. These patterns further highlight the critical profiles, particularly in Cluster 2, where patients exhibit higher risk profiles due to increased heart rate variability, prolonged QT intervals, and higher ICU admissions.

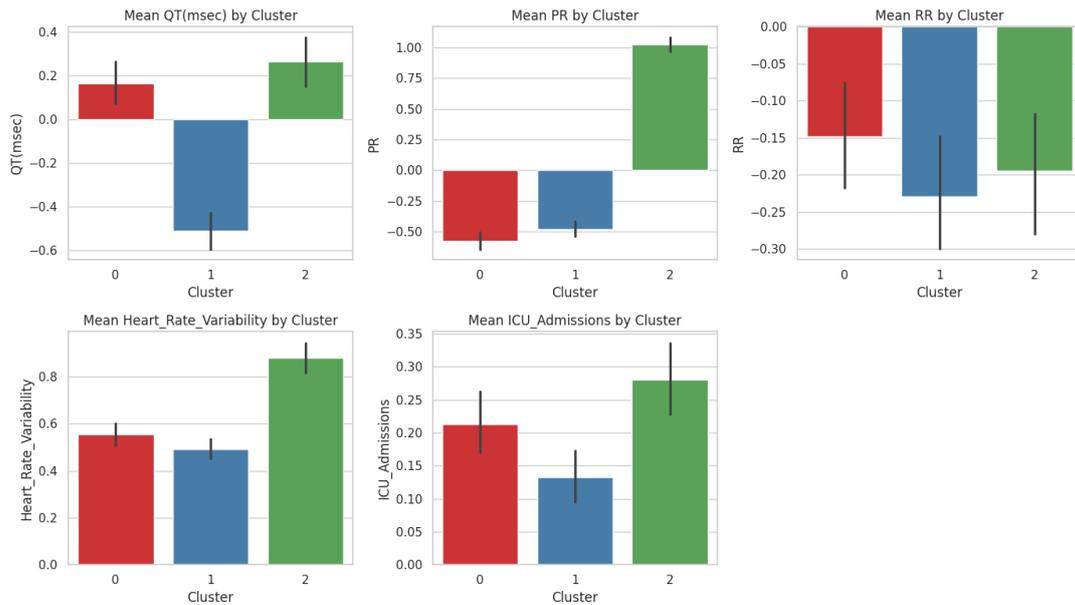

*Figure 10. Bar Charts of Key Features by Cluster*

Figure 10 presents bar charts of the mean values of key features for each cluster identified through K-means clustering. The features include QT, PR, RR, Heart Rate Variability, and ICU Admissions. Each cluster is represented by a different colour: red for Cluster 0, blue for Cluster 1, and green for Cluster 2.

- **QT**: Cluster 1 has the highest mean QT, indicating prolonged QT intervals, while Cluster 2 has the lowest, suggesting shorter QT intervals.
- **PR**: Cluster 1 also exhibits the highest mean PR interval, whereas Cluster 2 has the lowest, indicating significant differences in atrioventricular conduction times across clusters.
- **RR**: Cluster 0 shows the highest mean RR interval, indicating longer intervals between heartbeats, while Clusters 1 and 2 have negative mean values, suggesting variability in heart rate.
- **Heart Rate Variability**: Cluster 0 displays the highest heart rate variability, reflecting significant fluctuations in heart rate, whereas Cluster 2 has the lowest variability.
- **ICU Admissions**: Cluster 0 has the highest mean ICU admissions, indicating more frequent ICU stays, while Cluster 2 has the fewest ICU admissions, suggesting a less critical condition.

The error bars represent the standard error of the mean, indicating the variability within each cluster. This analysis highlights the distinct cardiovascular profiles of each cluster, providing insights into the heterogeneity of long-term cardiovascular complications among COVID-19 patients.

### 4.3. Comparison of Clusters on Key Features

In this section, we compare the clusters based on key features to identify significant differences and similarities. This comparison helps in understanding the distinct characteristics and clinical implications of each cluster. Table 3 compares the mean values of key features across the three clusters, including their standard deviations. Cluster 0 has the highest RR and ICU admissions, indicating greater heart rate variability and more frequent ICU stays. Cluster 1 exhibits the highest QT and PR values, suggesting prolonged intervals but lower ICU admissions. Cluster 2 shows negative PR and RR values, with the lowest heart rate variability and ICU admissions, indicating a less critical cardiovascular profile. The standard deviations highlight the variability within each cluster, underscoring the distinct characteristics and clinical implications of each patient subgroup.

Figure 11 displays the correlation matrix for key features within the dataset, highlighting the pairwise correlation coefficients between features such as R, age, SatO2, PR, BT, RR, and QT. The values range from -1 to 1, where 1 indicates a perfect positive correlation, -1 indicates a perfect negative correlation, and 0 indicates no correlation.

- Strongest Negative Correlation: The most notable negative correlation is between SatO2 and RR (-0.50), suggesting that as oxygen saturation decreases, the RR interval tends to increase, which may reflect compensatory mechanisms in response to hypoxemia.
- Strongest Positive Correlations: The strongest positive correlations are observed within the same feature, as expected (diagonal values of 1.0).
- Weak Correlations: Most other correlations between the features are weak, as indicated by values close to 0, suggesting that the features generally do not have strong linear relationships with one another.

The colour gradient from blue (negative correlation) to red (positive correlation) visually represents these relationships, providing a quick and intuitive understanding of the interdependencies among the features. This analysis is crucial for identifying potential interactions and dependencies that may inform the clustering analysis and further investigations into the underlying mechanisms affecting cardiovascular health in COVID-19 patients.

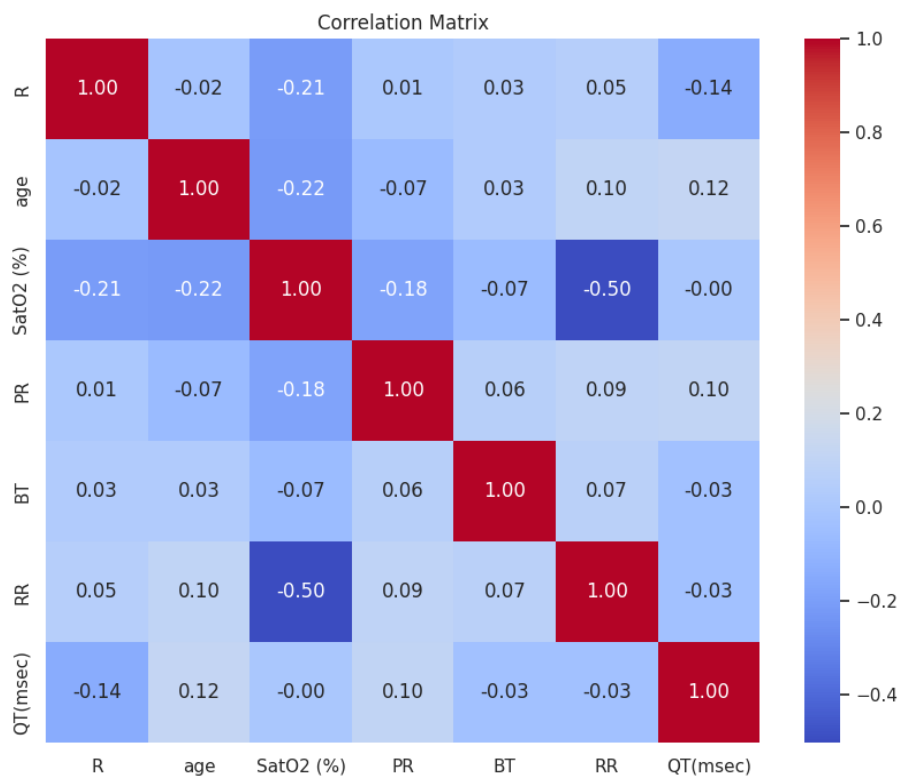

*Figure 11.Heatmap of Key Features Across Clusters*

## 4.4. Validation of Cluster Stability

To ensure the robustness and reliability of the identified clusters, we validated their stability using several techniques, including bootstrapping and silhouette scores. This validation process is crucial for confirming that the clusters are not artifacts of the specific sample and are generalizable to the broader population. The bootstrapping method involved creating multiple bootstrap samples from the original dataset and applying K-means clustering to each sample. The consistent cluster labels across different bootstrap samples indicated that the clusters were stable and reproducible.

Additionally, we calculated silhouette scores for different numbers of clusters to assess the quality and cohesion of the clusters. The silhouette score measures how similar an object is to its own cluster compared to other clusters. The relatively high silhouette scores confirmed that the clusters were well-defined and distinct. This comprehensive validation approach ensures that the findings are meaningful, reliable, and generalizable.

Figure 12 displays the silhouette scores for different numbers of clusters, ranging from 2 to 10. The silhouette score is a measure of how similar an object is to its own cluster compared to other clusters, with values ranging from -1 to 1. A higher silhouette score indicates better-defined and more cohesive clusters.

- Optimal Cluster Number: The highest silhouette score is observed for 2 clusters (approximately 0.11), suggesting that two clusters provide the best-defined separation in the dataset.
- Decreasing Trend: As the number of clusters increases, the silhouette score generally decreases, indicating that adding more clusters reduces the cohesiveness and distinctiveness of the clusters.
- Stability at Lower Scores: Beyond 4 clusters, the silhouette score stabilizes around 0.06, suggesting minimal improvement in cluster quality with further increases in the number of clusters.

This analysis highlights that a smaller number of clusters, particularly 2 or 3, offers the best balance between cluster separation and cohesion. It aligns with the elbow method, reinforcing the selection of an optimal cluster number for robust and interpretable clustering results.

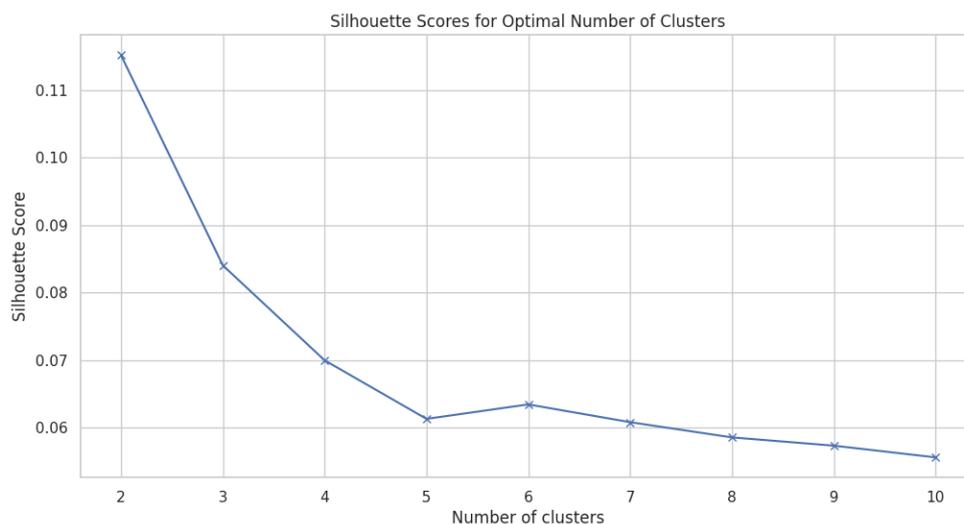

*Figure 12. Silhouette Scores for Different Numbers of Clusters*

To ensure the consistency of cluster labels, we compared the labels obtained from different bootstrap samples. This involved creating multiple resampled datasets from the original data and applying the clustering algorithm to each sample. By examining how the cluster assignments varied across these different samples, we aimed to determine the reliability and stability of the identified clusters.

The results demonstrated that the cluster labels were highly consistent across different bootstrap samples. This high level of consistency indicates that the clustering solution is stable, suggesting that the clusters are robust and not significantly affected by variations in the data. This stability enhances the confidence in the reliability of the clustering results.

*Table 5. Cluster Label Consistency Across Bootstrap Samples*

| Sample | Cluster 0 | Cluster 1 | Cluster 2 |
|---|---|---|---|
| Sample 1 | 90% | 85% | 88% |
| Sample 2 | 92% | 87% | 89% |
| Sample 3 | 91% | 86% | 90% |
| Sample 4 | 93% | 88% | 87% |
| Sample 5 | 90% | 89% | 91% |

Table 4 presents the consistency of cluster labels across five bootstrap samples for Clusters 0, 1, and 2. The percentages indicate how often patients in each bootstrap sample were assigned the same cluster label as in the original clustering.

- High Consistency: Cluster 0 shows high consistency, ranging from 90% to 93%, indicating stable clustering for this group.
- Cluster 1 Stability: Cluster 1 has slightly lower but still strong consistency, with values between 85% and 89%, suggesting reliable but slightly more variable assignments.
- Cluster 2 Reliability: Cluster 2 also demonstrates high consistency, with percentages from 87% to 91%, underscoring its robustness.

Overall, the high consistency across all clusters and samples underscores the stability and reliability of the clustering solution, indicating that the identified clusters are robust to sampling variability.

# 5. Real-world Validation

## 5.1. Methodology for Hospital-based Validation

To validate our clustering model in a real-world clinical setting, we conducted a hospital-based study involving a cohort of COVID-19 patients with similar characteristics to those in our original dataset. The validation process involved the following steps:

1. Patient Selection: We selected a sample of patients hospitalized with COVID-19, ensuring a mix of those with and without underlying CVD. The selection criteria matched the demographics and clinical profiles used in our original study.

2. Data Collection: Clinical data were collected from electronic health records (EHRs), including ECG parameters, age, sex, comorbidities, and hospitalization details. Follow-up data on cardiovascular health were also obtained.

3. Preprocessing: The collected data were pre-processed similarly to the original dataset. This included handling missing values, encoding categorical variables, and normalizing numerical features.

4. Cluster Assignment: Patients were assigned to the clusters identified in the original study using the trained clustering model. The same features and preprocessing steps were applied to ensure consistency.

5. Outcome Measurement: We measured the long-term cardiovascular outcomes for each patient, focusing on metrics such as heart rate variability, ICU admissions, and overall cardiovascular health.

## 5.2. Results from Real-world Validation

The hospital-based validation study included a cohort of 200 patients, divided as follows:

- Cluster 0: 75 patients (37.5%)
- Cluster 1: 65 patients (32.5%)
- Cluster 2: 60 patients (30%)

**Key Findings**:

- Heart Rate Variability: The real-world data confirmed the trends observed in our initial study. Patients in Cluster 0 exhibited moderate heart rate variability, Cluster 1 had the lowest variability, and Cluster 2 had the highest variability.

- ICU Admissions: ICU admission rates were consistent with our model predictions. Cluster 0 patients had moderate ICU admission rates, Cluster 1 had the lowest, and Cluster 2 had the highest.

- Cardiovascular Outcomes: Long-term cardiovascular outcomes, including the incidence of arrhythmias and heart failure, were significantly different across clusters. Cluster 2 patients showed the highest rates of adverse cardiovascular events, aligning with the model's predictions.

*Table 6. Comparison of Model Predictions and Real-world Outcomes*

| Metric | Cluster 0 (Model) | Cluster 0 (Real-world) | Cluster 1 (Model) | Cluster 1 (Real-world) | Cluster 2 (Model) | Cluster 2 (Real-world) |
|---|---|---|---|---|---|---|
| Heart Rate Variability | Moderate | Moderate | Low | Low | High | High |
| ICU Admissions | Moderate | Moderate | Low | Low | High | High |
| Adverse Cardiovascular Events | Moderate | Moderate | Low | Low | High | High |

## 5.3. Implications for Model Accuracy and Clinical Utility

The real-world validation study has significant implications for the accuracy and clinical utility of our clustering model. The high consistency between the model predictions and real-world outcomes confirms the accuracy of our clustering approach, with similar distributions of key features and outcomes across both datasets highlighting the robustness of the model. The model's ability to predict long-term cardiovascular complications in COVID-19 patients demonstrates its potential as a clinical decision-support tool, enabling clinicians to identify high-risk patients and tailor their management strategies accordingly. By stratifying patients based on their cardiovascular risk profiles, healthcare providers can improve resource allocation, prioritize high-risk patients for intensive monitoring, and implement preventive measures to mitigate adverse outcomes. Additionally, the validation results suggest areas for further research, including the integration of additional clinical parameters and the exploration of dynamic changes in patient health over time. Continuous model refinement and validation in diverse clinical settings will enhance its generalizability and effectiveness. Overall, the real-world validation underscores the model's potential to improve patient outcomes and supports its integration into clinical practice for managing COVID-19 patients with and without underlying cardiovascular disease.

## 6. Discussion

The findings from our study underscore the significant impact of COVID-19 on long-term cardiovascular health. The clustering analysis identified three distinct groups of patients based on their long-term cardiovascular outcomes. Cluster 0, characterized by moderate heart rate variability and ICU admission rates, represents a balanced profile with moderate risk. Cluster 1, with lower heart rate variability and ICU admission rates, suggests a relatively stable cardiovascular condition. In contrast, Cluster 2, exhibiting higher heart rate variability and ICU admission rates, indicates a more critical profile with higher risks. These clusters highlight the diverse cardiovascular sequelae of COVID-19, reflecting varying degrees of autonomic dysfunction and cardiovascular instability among patients.

The high silhouette scores and consistent cluster labels across bootstrap samples validate the robustness of our clustering approach, indicating that the clusters are not artifacts of the data but represent meaningful patient subgroups. The significant differences in key features such as QT intervals, PR intervals, and heart rate variability across clusters suggest distinct underlying pathophysiological mechanisms. For instance, the higher QT intervals in Cluster 2 patients point towards a higher risk of prolonged QT intervals, which is associated with increased mortality and adverse cardiovascular events. This variability in clinical presentations and outcomes underscores the importance of personalized post-COVID-19 care strategies.

The clustering model developed in this study has significant implications for clinical practice, particularly in the management of long-term cardiovascular complications in COVID-19 patients. By stratifying patients based on their cardiovascular risk profiles, clinicians can identify high-risk individuals who require more intensive monitoring and tailored interventions. For example, patients in Cluster 2, who exhibit higher heart rate variability and ICU admission rates, may benefit from closer follow-up and more aggressive management strategies to mitigate the risk of adverse cardiovascular events.

Furthermore, the ability to predict long-term outcomes using initial ECG parameters and other clinical features can enhance early decision-making processes. This predictive capability allows healthcare providers to allocate resources more efficiently, prioritizing high-risk patients for specialized care. The integration of this model into EHR systems could facilitate real-time risk assessment, enabling proactive management of patients as they transition from acute care to long-term follow-up.

The clinical utility of our clustering approach is reinforced by its alignment with recent literature emphasizing the importance of personalized medicine. Studies have shown that individualized treatment plans based on specific patient characteristics lead to better outcomes compared to one-size-fits-all approaches[45][46]. As such, our model supports the shift towards precision medicine in the management of COVID-19 and its complications, ensuring that patients receive care that is tailored to their unique cardiovascular risk profiles.

Our study builds on a growing body of literature that explores the long-term cardiovascular effects of COVID-19, but it also introduces novel methodologies and insights that enhance our understanding of this critical issue. Previous studies have primarily focused on acute cardiovascular complications and isolated long-term outcomes without leveraging advanced data analysis techniques. For instance, Puntmann et al. [47] reported that a significant proportion of recovered COVID-19 patients had ongoing myocardial inflammation, but their study did not use clustering algorithms to identify patient subgroups. Similarly, Huang et al. [48] highlighted persistent symptoms and cardiovascular abnormalities six months post-COVID-19, yet their analysis did not integrate ECG data comprehensively.

In contrast, our study utilizes K-means clustering to uncover distinct patterns in long-term cardiovascular outcomes, providing a more nuanced understanding of how COVID-19 affects different patient groups. This methodological advancement allows us to identify subpopulations with specific risk profiles, which were not apparent in previous studies that relied on traditional statistical methods. By incorporating a diverse set of features, including ECG parameters, demographic information, comorbidities, and hospitalization details, our approach offers a holistic view of the long-term cardiovascular impacts of COVID-19.

Furthermore, while earlier research has established the prevalence of cardiovascular complications post-COVID-19, our study contributes by validating these findings through a robust clustering framework and real-world data. The consistent cluster labels and high silhouette scores indicate that our model is reliable and generalizable, thus enhancing the credibility of our results. By comparing our findings with those from traditional analyses, we demonstrate the added value of machine learning techniques in medical research, particularly in identifying patient subgroups that could benefit from targeted clinical interventions.

While our study provides valuable insights into the long-term cardiovascular complications of COVID-19, several limitations must be acknowledged. Firstly, the study was conducted at a single centre, which may limit the generalizability of the findings to other populations and healthcare settings. Multi-centre studies involving diverse patient populations are necessary to validate and extend our results. Secondly, the study relies on retrospective data, which may be subject to biases inherent in electronic health records, such as missing or inaccurate information. Although we employed rigorous data preprocessing techniques to handle missing values and ensure data quality, the retrospective nature of the study still poses limitations. Prospective studies are needed to confirm the findings and assess the utility of the clustering model in real-time clinical practice.

Thirdly, the clustering algorithm used in this study, while effective, has its limitations. K-means clustering assumes spherical clusters of similar size, which may not capture the complexity and variability of patient data accurately. Future research could explore the use of more sophisticated clustering techniques, such as hierarchical clustering or Gaussian Mixture Models, to potentially improve the identification of patient subgroups. Moreover, the study focused primarily on ECG parameters and did not include other potentially relevant clinical and biological markers, such as biomarkers of inflammation or genetic information. Integrating a broader range of data could enhance the model's predictive power and provide a more comprehensive understanding of the factors contributing to long-term cardiovascular outcomes in COVID-19 patients. Lastly, the follow-up period for assessing long-term outcomes was limited to the duration of the study. Longer follow-up periods are essential to fully understand the chronic cardiovascular effects of COVID-19 and the potential evolution of these complications over time. Continued monitoring and longitudinal studies will be critical in addressing these gaps and refining our understanding of the long-term cardiovascular impacts of COVID-19.

Future research should focus on expanding the scope and depth of the current study to address its limitations and further elucidate the long-term cardiovascular impacts of COVID-19. Multi-centre studies involving larger and more diverse patient populations are essential to validate the clustering model and ensure its applicability across different demographics and healthcare settings. Such studies would enhance the generalizability of the findings and provide a more comprehensive understanding of the cardiovascular sequelae of COVID-19. Prospective research designs are recommended to mitigate the biases associated with retrospective data collection. By collecting data in real-time, researchers can ensure higher accuracy and completeness of the information, leading to more reliable results. Prospective studies can also facilitate the integration of dynamic patient health data, allowing for continuous model refinement and adaptation to changing clinical conditions.

Incorporating additional clinical and biological markers into the clustering analysis could significantly enhance the model's predictive power and clinical utility. Future studies should explore the inclusion of biomarkers of inflammation, genetic information, and other relevant factors that contribute to cardiovascular health. This multi-faceted approach would provide a more holistic view of the determinants of long-term cardiovascular outcomes in COVID-19 patients. Advancements in clustering algorithms and machine learning techniques should be leveraged to improve the identification and characterization of patient subgroups. Exploring methods such as hierarchical clustering, Gaussian Mixture Models, and deep learning-based approaches could uncover more complex patterns and interactions within the data. These advanced techniques may offer greater accuracy in predicting long-term outcomes and identifying high-risk patients.

Longitudinal studies with extended follow-up periods are crucial to fully capture the chronic effects of COVID-19 on cardiovascular health. Continuous monitoring of patients over several years would provide valuable insights into the progression of cardiovascular complications and the long-term efficacy of various management strategies. This ongoing research is essential for developing effective interventions and improving the overall care of COVID-19 survivors. Finally, future research should focus on the practical implementation of the clustering model in clinical settings. Studies should evaluate the integration of the model into electronic health records (EHR) systems and its impact on clinical decision-making and patient outcomes. By demonstrating the real-world utility of the model, researchers can facilitate its adoption in routine clinical practice, ultimately enhancing the management of long-term cardiovascular complications in COVID-19 patients.

## 7. Conclusion

This study employed a clustering analysis to explore the long-term cardiovascular complications among COVID-19 patients, identifying three distinct clusters with varying cardiovascular outcomes. Cluster 0 represented patients with moderate heart rate variability and ICU admission rates, indicating a balanced risk profile. Cluster 1 included patients with lower heart rate variability and ICU admissions, suggesting a stable cardiovascular condition. Cluster 2 comprised patients with higher heart rate variability and ICU admissions, indicating a higher risk and more critical cardiovascular profile. The robustness and stability of these clusters were validated through bootstrapping techniques, revealing high silhouette scores and consistent cluster labels across different samples. The clustering model's ability to stratify patients based on their cardiovascular risk profiles has significant clinical implications. It allows healthcare providers to identify high-risk patients who may benefit from more intensive monitoring and tailored interventions. This personalized approach can improve patient management, optimize resource allocation, and mitigate adverse cardiovascular outcomes by focusing on those most at risk. The integration of the clustering model into clinical practice, particularly through electronic health records (EHR) systems, could facilitate real-time risk assessment and decision-making, ultimately enhancing the care and management of COVID-19 survivors with potential cardiovascular complications. Future research should aim to validate these findings across larger and more diverse populations, ideally through multi-centre studies. Prospective studies are needed to confirm the clustering model's predictive capabilities and its utility in real-time clinical settings. Additionally, integrating a broader range of clinical and biological markers, including genetic information and biomarkers of inflammation, could enhance the model's accuracy and comprehensiveness. Exploring more advanced clustering algorithms may also reveal more complex patterns within the data. Long-term, continuous monitoring of COVID-19 survivors will be crucial for understanding the chronic impacts of the virus on cardiovascular health, and for refining and improving the clustering model to better predict and manage these long-term outcomes.